\documentclass[usenatbib]{mn2e}
\usepackage{natbib,psfig}

\catcode`\@=11
\def\gsim{\ifmmode{\mathrel{\mathpalette\@versim>}}
    \else{$\mathrel{\mathpalette\@versim>}$}\fi}
\def\lsim{\ifmmode{\mathrel{\mathpalette\@versim<}}
    \else{$\mathrel{\mathpalette\@versim<}$}\fi}
\def\@versim#1#2{\lower 2.9truept \vbox{\baselineskip 0pt \lineskip
    0.5truept \ialign{$\m@th#1\hfil##\hfil$\crcr#2\crcr\sim\crcr}}}
\catcode`\@=12

\newcommand{\beq}{\begin{equation}}
\newcommand{\eeq}{\end{equation}}
\newcommand{\az}{{a_0}}
\newcommand{\phiN}{\phi^{\rm N}}
\newcommand{\phiNtot}{\phi^{\rm N}_{\rm tot}}
\newcommand{\rhotot}{\rho_{\rm tot}}
\newcommand{\gstarN}{g_*^{\rm N}}
\newcommand{\gv}{{\bf g}}
\newcommand{\gvN}{{\bf g}^{\rm N}}
\newcommand{\Sv}{{\bf S}}
\newcommand{\rhos}{\rho_*}
\newcommand{\Mstar}{M_*}
\newcommand{\Mstarten}{M_{*,10}}
\newcommand{\tstar}{t_*}
\newcommand{\tg}{t_{\rm g}}
\newcommand{\Tz}{T_0}

\newcommand{\Omegabz}{\Omega_{\rm b,0}}
\newcommand{\Omegab}{\Omega_{\rm b}}
\newcommand{\Omegabdot}{\dot{\Omega}_{\rm b}}
\newcommand{\Omegastar}{\Omega_*}
\newcommand{\MDM}{M_{\rm DM}}
\newcommand{\rstar}{r_*}

\newcommand{\gammadm}{\gamma_{\rm dm}}
\newcommand{\vstar}{v_*}

\newcommand{\Npart}{N_{\rm part}}

\newcommand{\tfricM}{t_{\rm fric}^{\rm M}}
\newcommand{\tfricN}{t_{\rm fric}^{\rm N}}
\newcommand{\tfric}{t_{\rm fric}}

\newcommand{\rhodm}{\rho_{\rm dm}}

\newcommand{\vc}{v_{\rm c}}

\newcommand{\bey}{\begin{eqnarray}}
\newcommand{\eey}{\end{eqnarray}}
\newcommand{\Myr}{\, {\rm Myr} }

\newcommand{\kpc}{\, {\rm kpc} }
\newcommand{\Msun}{M_\odot}

\newcommand{\kms}{\, {\rm km \, s}^{-1} }

\newcommand{\xv}{{\bf x}}

\newcommand{\mathR}{{\mathcal R}}

\def\xv{{\bf x}}

\def\varphib{\varphi_{\rm b}}
\def\varphibdot{\dot{\varphi}_{\rm b}}

\def\xv{{\bf x}}

\def\rhob{\rho_{\rm b}}
\def\Mb{M_{\rm b}}
   \title[Dynamical friction in MOND]
         {Dynamical friction in Modified Newtonian Dynamics}

\author[]{}

   \author[Nipoti et al.]
          {Carlo Nipoti$^1$, 
           Luca Ciotti$^1$,  James Binney$^2$, and Pasquale Londrillo$^3$.
           \\ $^1$Astronomy Department, University of Bologna, 
                       via Ranzani 1, 40127 Bologna, Italy
	   \\ $^2$Rudolf Peierls Centre for Theoretical Physics, University
			of Oxford, 1 Keble Road, Oxford OX1 3NP, UK
           \\ $^3$INAF-Bologna Astronomical Observatory, 
                       via Ranzani 1, 40127 Bologna, Italy
           }

\date{Accepted 2008 March 3. Received 2008 February 29; in original form 2008 February 8}
\pubyear{2008}

\begin{document} 
\maketitle

\begin{abstract}

  We have tested a previous analytical estimate of the dynamical
  friction timescale in Modified Newtonian Dynamics (MOND) with fully
  non-linear N-body simulations. The simulations confirm that the
  dynamical friction timescale is significantly shorter in MOND than
  in equivalent Newtonian systems, i.e. systems with the same
  phase-space distribution of baryons and additional dark matter. An
  apparent conflict between this result and the long timescales
  determined for bars to slow and mergers to be completed in previous
  N-body simulations of MOND systems is explained. The confirmation of
  the short dynamical-friction timescale in MOND underlines the
  challenge that the Fornax dwarf spheroidal poses to the viability of
  MOND.

\end{abstract}


\section{Introduction}
\label{secint}

Numerous observations indicate that galaxies and clusters of galaxies
have gravitational fields that are stronger at large radii than the
standard theory of gravity predicts from the conjecture that light is
a fair tracer of mass. The standard interpretation of this phenomenon
is that $\sim\frac45$ of the mass in the Universe is contributed by
particles that are dark because they do not interact
electromagnetically. The alternative hypothesis, that the standard
theory of gravity fails at low accelerations, was advanced by
\cite{Mil83}.  Subsequently \cite{BekM84} proposed a modification of
Poisson's equation that put the proposal on a quantitative basis.
Recently interest in Modified Newtonian Dynamics (MOND) has increased
with a growing awareness that relativistically covariant theories are
possible that reduce in the non-relativistic limit to the
Bekenstein-Milgrom theory \citep{Bek04}.

One way to rule out MOND would be to detect the particles that are
supposed to contribute most of the Universe's mass, and several
experiments are endeavouring to do this under the assumption that the
particles have weak interactions. Another way to rule out MOND would
be to show that it is inconsistent with observations of the growth of
cosmic structure \citep[e.g.][]{Pedro} or the dynamics and evolution of
galaxies.

In standard gravity, dynamical friction (DF) is thought to play a
major role in several astrophysical contexts
\citep[e.g.][\S8.1]{BT08}.  Ciotti \& Binney (2004; hereafter CB04)
showed that DF is a more potent phenomenon in MOND than in standard
gravity, and argued that it would cause the globular clusters of dwarf
spheroidal galaxies to spiral to the centres of their hosts within a
dynamical time. However, the results in CB04 were obtained under
rather restrictive assumptions and by a non-standard argument that
involves the fluctuation-dissipation theorem \citep[see][]{BekM92}. In
particular, CB04 gave results for plane-parallel systems that are in
the ``deep-MOND'' regime.  Given the prediction of CB04 that DF is
much more powerful in MOND than in Newtonian gravity, it is important
to test their results with a different approach and in less
specialized contexts.

In this paper we use N-body experiments to explore DF in regimes in
which departures from Newtonian gravity are of varying magnitude.  The
paper is organized as follows. In Section~2 we outline our methodology
explain our choice of problem. Section~3 gives details of the code and
the initial conditions of the simulations. Section 4 presents the
results and Section~5 relates them to other recent work. The
conclusions and their astronomical implications are summarized in
Section~6.

\section{Methodology}

Two fundamental features of MOND preclude a direct extension of the
standard Chandrasekhar--Spitzer derivation of DF \citep{Cha42,Spi87}:
in MOND (i) forces are not additive and (ii) all two-body orbits are
bound, so individual encounters do not have a finite duration. On the
other hand, there is no fundamental obstacle to direct simulations of
DF in MOND: the drag experienced by a massive object as it moves
through a swarm of less massive objects can be measured in an N-body
simulation.

In the classical approach to DF one imagines the background swarm to be
homogeneous and perpetrates the Jeans swindle to neglect the mean
gravitational field of the swarm. Even in Newtonian gravity this step is of
questionable validity, and it is inadmissible in MOND because (i) the
long-range nature of two-particle interactions in MOND implies that distant
encounters are liable to dominate (in the Newtonian case they nearly do) and
(ii) the non-linearity of the Bekenstein--Milgrom field equation implies
that the impact of an encounter depends on the nature of the mean field.
Therefore, any attempt to extend to MOND the classical approach to DF leads
to the consideration of the drag that a massive body experiences as it moves
through a self-gravitating system of finite size.

In order to compare the action of DF in MOND and Newtonian systems as
cleanly as possible, we compare DF in MOND with DF in the {\it equivalent
Newtonian system} (ENS); that is the Newtonian system in which the visible
matter has exactly the same phase-space distribution as in the MOND system
\citep{Mil01,Nip07b,NipLC07c}.  In the ENS the visible matter is
enveloped in a dark-matter halo that  contributes to the DF experienced
by a massive body.

There are two astrophysically important problems for which DF is
crucial: (i) the inward spiralling of a small system that has fallen
into a larger one \citep[e.g.][]{TreOS75,BonV87,HernquistW,AreB07},
and (ii) the slowing of a massive bar by the material in which it is
embedded \citep[e.g.][and references therein]{Wei85,Deb98,Sel06}. Our
MOND-enabled N-body code \citep[N-MODY;][]{CioLN06,NipLC07a,LonN08} is
better adapted for problem (ii) because it uses a polar grid and
therefore has higher resolution at small radii than at large. So we
study the effect that MOND has on the rate at which the rotation of a
rigid bar is slowed by DF.

\begin{table}

 \flushleft{

%\begin{minipage}{80mm}

  \caption{Parameters of the simulations.}

  \begin{tabular}{llccccc}
     Name    & Gravity &  $\gamma$ &$\MDM/\Mstar$ & $\kappa$ & $\Omegabz/\Omegastar$ & $\Npart$ \\
     ~~(1) & ~~~(2) & (3) & (4) & (5) &  (6) &  (7)\\ 
%    [10pt]
\hline
%      grav    gam   mdm  kap     omegabz Ntot  NDM 
M00    & MOND   & 0 &~0   & ~1    & 0.61 & $~8$ \\
E00    & Newton & 0 &~6.6 & ~1    & 0.61 & $16$ \\
M01    & MOND   & 0 &~0   & 0.1   & 1.06 & $~8$ \\
E01    & Newton & 0 &~23.1& 0.1   & 1.06 & $16$ \\
M02    & MOND   & 0 &~0   & 0.01  & 1.88 & $~8$ \\
E02    & Newton & 0 &~75.3& 0.01  & 1.88 & $16$ \\
M10    & MOND   & 1 &~0   & ~1    & 0.74 & $~8$ \\
E10    & Newton & 1 &~7.0 & ~1    & 0.74 & $16$ \\
M11    & MOND   & 1 &~0   & 0.1   & 1.27 & $~8$ \\
E11    & Newton & 1 &~24.4& 0.1   & 1.27 & $16$ \\
M12    & MOND   & 1 &~0   & 0.01  & 2.26 & $~8$ \\
E12    & Newton & 1 &~79.2& 0.01  & 2.26 & $16$ \\
\hline

\end{tabular}

}

\medskip

\flushleft{(1): name of the simulation. (2): gravity law. (3) inner
logarithmic slope of the stellar density distribution
[equation~(\ref{eqrhostar})]. (4): dark matter to baryonic mass
ratio. (5): internal acceleration ratio $\kappa\equiv
G\Mstar/(\az\rstar^2)$. (6): initial angular frequency of the bar, in
units of $\Omegastar=1/\tstar$.  (7): total number of particles in
units of $10^6$.}
%\end{minipage}
\end{table}

\section{The numerical simulations}
\label{secmet}

With \cite{BekM84} we assume that Poisson's equation $\nabla^2\phiN=4\pi
G\rho$ should be replaced by the non-relativistic field equation
 \begin{equation}
\nabla\cdot\left[\mu\left({\Vert\nabla\phi\Vert\over\az}\right)
\nabla\phi\right] = 4\pi G \rho,
\label{eqMOND}
\end{equation} 
 where $\Vert ...\Vert$ is the standard Euclidean norm, and $\phi$ is the
gravitational potential for MOND; the gravitational acceleration is
$\gv=-\nabla\phi$ just as the Newtonian acceleration is $\gvN=-\nabla\phiN$.
For a system of finite mass, $\nabla\phi\to 0$ as $\Vert\xv\Vert\to\infty$.
The function $\mu(y)$ is constrained by the theory only to the extent that
it must run smoothly from $\mu(y)\sim y$ at $y\ll 1$ (the so-called
deep-MOND regime) to $\mu(y)\sim 1$ at $y\gg 1$ (the Newtonian regime), with
the transition taking place at $y\approx 1$, i.e., when $\Vert\nabla\phi\Vert$
is of order the characteristic acceleration $\az
\simeq 1.2 \times 10^{-10} {\rm m}\,{\rm s}^{-2}$. In the present work
we adopt $\mu(y)=y/\sqrt{1+y^2}$ \citep{Mil83}.

From Poisson's equation and equation~(\ref{eqMOND}) it follows that
the MOND and Newtonian gravitational accelerations are related by
${\mu}(g/\az) \, \gv = \gvN +\Sv$, where $g\equiv\Vert\gv\Vert$, and
$\Sv$ is a solenoidal field dependent on the specific $\rho$
considered. Since in general $\Sv\ne0$, standard Poisson solvers
cannot be used to develop MOND N-body codes; equation~(\ref{eqMOND})
must be solved at each time step (Brada \& Milgrom 1999, Nipoti et
al. 2007a).  

\subsection{The code}

N-MODY is a parallel, three-dimensional particle-mesh code that can be
used to run either MOND or Newtonian simulations \citep[Nipoti et
al. 2007a,][]{LonN08}. In the present study the spherical grid has 128
radial nodes, 64 nodes in colatitude $\theta$ and 128 nodes in azimuth
$\phi$, and the total number of particles is in the range
$\Npart=8\times10^6-1.6\times10^7$. We verified with convergence
experiments that these numbers of particles and grid points are
sufficient to exclude that our results are significantly affected by
discreteness effects.

In previous papers \citep[][Nipoti et al. 2007ac]{CioNL07} we have
used N-MODY to demonstrate significant differences in the operation of
violent relaxation in systems in which MOND is important and in their
Newtonian equivalents. We refer readers to these papers and to
\cite{LonN08} for details of the code and its tests.

\subsection{Initial conditions}
\label{secini}

The baryonic component of the initial conditions of the
simulations is described by a spherical
$\gamma$-model \citep{Deh93,Tre94} with density distribution
\begin{equation}
\rhos (r)= {3-\gamma\over 4\pi}{\Mstar\rstar \over 
r^{\gamma} (\rstar+r)^{4-\gamma}}  ,
\label{eqrhostar}
\end{equation}
where $\Mstar$ is the total stellar mass, $\rstar$ is the scale
radius, $\gamma$ is the inner logarithmic slope, and we consider the
cases $\gamma=0$ and $\gamma=1$ \citep{Her90}.  For a given model, the
MOND potential is easily calculated, and to each MOND model with
potential $\phi$ corresponds an ENS with $\phiNtot=\phi$, thus having
a {\it total} density $\rhotot(r)=\nabla^2\phi(r) /4\pi G$. In
principle, such a distribution would have infinite mass, so we
truncate it at $r\sim10\rstar$.

The particles of the stellar component are distributed in phase-space
with the standard rejection technique, restricting for simplicity to
the fully isotropic case. The MOND distribution function is obtained
numerically with an Eddington inversion \citep[e.g.][\S4.3.1]{BT08}
\begin{equation}
f_M(E)={1\over \sqrt{8}\pi^2}{d\over dE}\int_E^{\infty}{d\rhos\over d\phi}
{d\phi\over\sqrt{\phi -E}},
\label{eqdf}
\end{equation}
where the upper integration limit reflects the far-field logarithmic
behaviour of the MOND potential.

In the equivalent Newtonian models, at variance with Nipoti et
al.~(2007c), we do not distinguish between dark and stellar particles,
but we consider a single component distributed according to the
numerical isotropic distribution function
\begin{equation}
f_N(E)={1\over \sqrt{8}\pi^2}{d\over dE}\int_E^{0}{d\rhotot\over d\phiNtot}
{d\phiNtot\over\sqrt{\phiNtot -E}}.
\label{eqdfN}
\end{equation}
Note that the upper limit of integration is due to the finite mass of
the system (a consequence of the density truncation at $10\rstar$).

The physical scales of the problem are introduced as follows.  First
we identify each MOND initial condition by fixing a value for the
dimensionless internal acceleration parameter $\kappa\equiv
G\Mstar/(\az\rstar^2)$, so that $\Mstar$ and $\rstar$ are not
independent quantities: in physical units, $\rstar\simeq3.4
\kappa^{-1/2}\Mstarten^{1/2}\kpc$, where $\Mstarten\equiv
\Mstar/10^{10}\Msun$.  The time and velocity units are
$\tstar=\sqrt{\rstar^3/G\Mstar} \simeq
29.7\kappa^{-3/4}\Mstarten^{1/4}\Myr$, and $\vstar = \rstar/\tstar
\simeq 112\kappa^{1/4}\Mstarten^{1/4}\kms$ (Nipoti et al.~2007a).  The
simulations are evolved up to $t=30\Tz$ (where $\Tz$ is the initial
rotation period of the bar, see Section~\ref{secbar}), with timestep
$\Delta t =0.003 \Tz$.

%%%%%%%%%%%%%%%%%%%% FIG 1
\begin{figure}
\centerline{ 
\psfig{figure=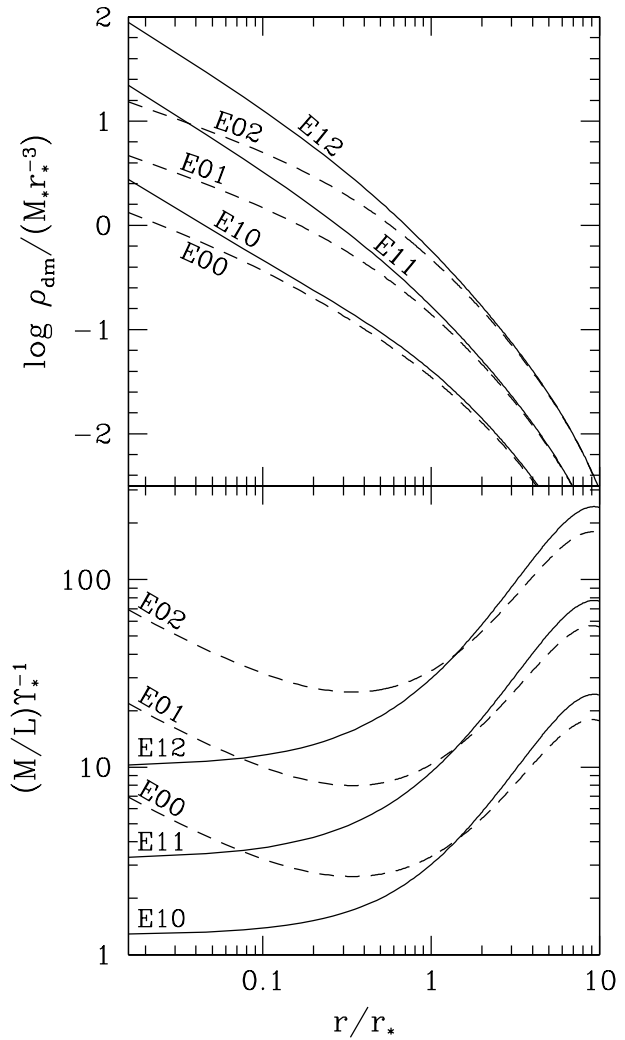,width=\hsize,angle=0,bbllx=20bp,bblly=720bp,bburx=350bp,bbury=170bp,clip=}} 
\caption{Dark-matter density $\rhodm\equiv\rhotot-\rhos$ (top
  panel) and total (dark-matter plus stars) mass-to-light ratio $M/L$
  (bottom panel) as functions of radius for the equivalent Newtonian
  models. $M/L$ is in units of the stellar mass-to-light ratio
  $\Upsilon_*$. The first digit after ``E'' is the value of $\gamma$
  while the second digit is $-\log_{10}(\kappa)$ (Table~1). Dashed and
  solid curves refer to models with $\gamma=0$ and $\gamma=1$,
  respectively.}
\label{figmlratio}
\end{figure}
%%%%%%%%%%%%%%%%%%%% FIG 2
\begin{figure*}
\centerline{ \psfig{file=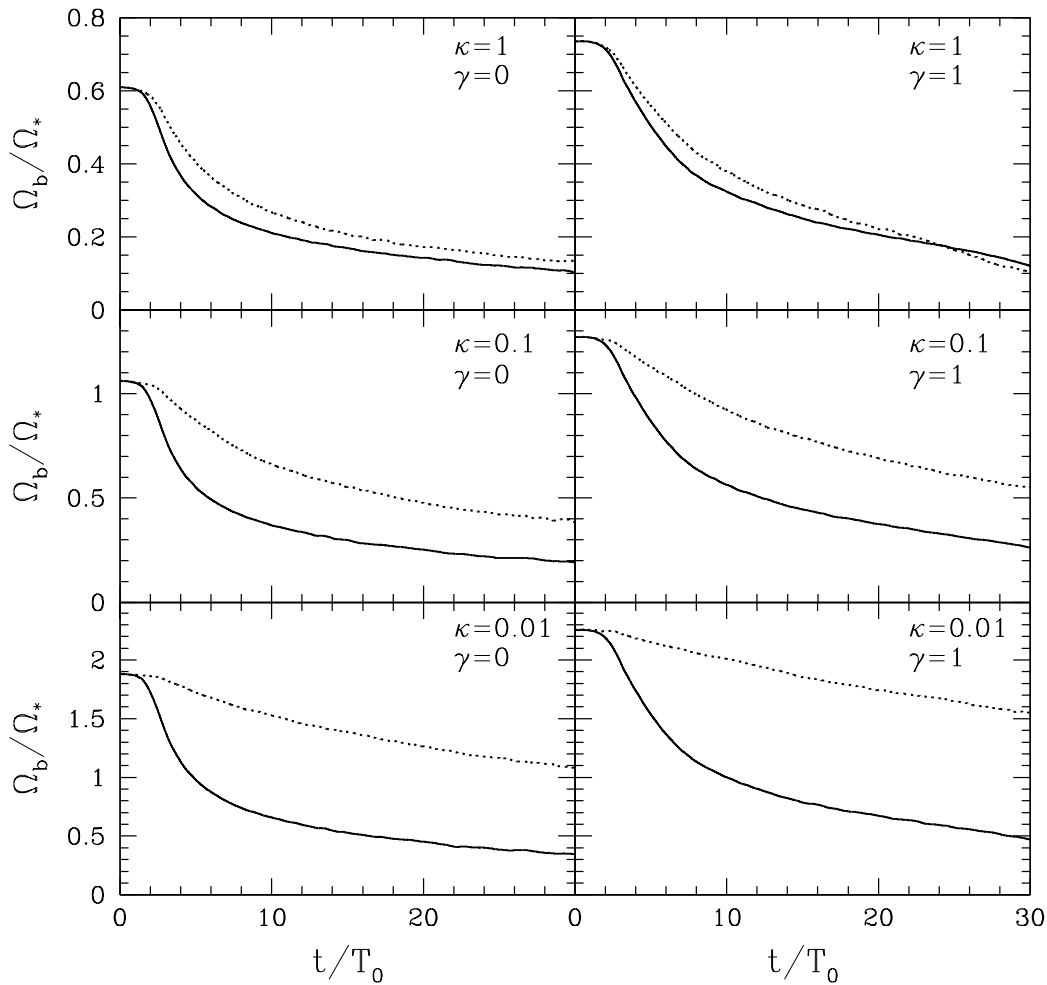,width=.8\hsize}}
\caption{Time evolution of the bar angular frequency in units of
  $\Omegastar=1/\tstar$ in MOND (solid curves) and equivalent
  Newtonian models with dark matter (dotted curves), for six pairs of
  models with different combinations of the parameters $\kappa$ and
  $\gamma$. Cuspy models are on the right. Smaller values of $\kappa$ correspond to deeper MOND
  regimes.  Time is normalized to the initial rotation period of the
  bar $\Tz$.}
\label{figtomega}
\end{figure*}
%%%%%%%%%%%%%%%%%%%%

We verified that both the MOND system and its ENS were in equilibrium
by running them without the bar for several dynamical times.  The dark
and luminous matter distributions of the considered ENSs are shown in
Fig.~\ref{figmlratio}, plotting, as functions of radius, their
dark-matter density $\rhodm\equiv\rhotot-\rhos$ (top panel) and their
total (dark-matter plus stars) mass-to-light ratio $M/L\equiv
\Upsilon_* \rhotot/\rhos$, where $\Upsilon_*$ is the stellar
mass-to-light ratio (bottom panel). All the considered ENSs are
dark-matter dominated, because the corresponding MOND cases have
internal accelerations $\lsim \az$.  Focusing on the central regions
of the ENSs, we note that, for fixed $\kappa$, models with cored
stellar profile ($\gamma=0$, dashed curves in Fig.~\ref{figmlratio})
have higher $M/L$ and shallower inner dark-matter profile than models
with cuspy stellar profile ($\gamma=1$, solid curves in
Fig.~\ref{figmlratio}). In particular, we have $\gammadm\sim 1$ when
$\gamma = 1$ and $\gammadm\sim 1/2$ when $\gamma = 0$, where
$\gammadm\equiv-\lim_{r\to 0} (d\ln \rhodm / d\ln r)$ is the inner
logarithmic slope of the dark-matter density distribution.

\subsection{Bar equation of motion and the warm-up of the gravitational field}
\label{secbar}

The density distribution of the rigid bar is a prolate ellipsoid of
density distribution
\begin{equation}
\rhob={\Mb\over \pi^{3/2}q b^3}\,{\rm e}^{-m^2/b^2},
\label{eqrhob}
\end{equation}
where in the Cartesian system co-rotating with the bar
\begin{equation}
m^2=x'^2+z'^2+{y'^2\over q^2},\quad q\geq 1.
\label{eqmsq}
\end{equation}
 In all the cases here presented we use $\Mb=0.05\Mstar$, $b=0.2\rstar$ and
$q=5$: with this choice, the length of the bar's semi-major axis is $\rstar$.
We adopted the distribution~(\ref{eqrhob}) because the Gaussian
dependence on $m$ produces a bar well defined spatially, but at the same
time it avoids a density discontinuity that could produce spurious behaviour
in the potential solver at the bar's edge.  In all the simulations the
bar rotates around the $z'$ axis.

The Lagrangian formulation can be used to derive the bar's equations of
motion in an inertial Cartesian frame $(x,y,z)$ such that $z=z'$: with
$\varphib$ the angle in the $(x,y)$ plane between the $x$ and $x'$ axes,
we have 
 \begin{equation}
I_{33}{d^2\varphib\over dt^2}=\int\rhob (m)\left(
                             y{\partial\phi\over\partial x}-
                             x{\partial\phi\over\partial y}\right)dV,
%=-\int \rhob (m){\partial \phi \over \partial \varphi} dV,
\label{eqphib}
\end{equation}
where 
\begin{equation}
m^2= (x\cos\varphib +y\sin\varphib)^2+z^2 +
     {(y\cos\varphib -x\sin\varphib)^2\over q^2}
\end{equation}
and from equation~(\ref{eqrhob})
\begin{equation}
I_{33}={4\pi q(1+q^2)\over 3}\int_0^{\infty}m^4\rhob(m)dm=
{\Mb b^2 (1+q^2)\over 2}.
\label{eqinertia}
\end{equation}

In order to start  the numerical simulations smoothly, we
take the density of the bar to be the product of equation~(\ref{eqrhob}) and
the function
 \begin{equation}
\alpha(t)\equiv \min\left({t^3\over \tg^3},1\right),
\end{equation}
 where the growth time is $\tg=3\Tz$ with $\Tz\equiv2\pi/\Omegabz$ and
 $\Omegabz$ is the initial angular velocity of the bar's frame.

As a consequence, when the bar mass is very small, the underlying
density distribution is only slightly affected by its gravitational
field, and the integral at the r.h.s of equation~(\ref{eqphib}) almost
vanishes.  Of course, in the MOND simulations, on the r.h.s. of
equation~(\ref{eqMOND}) we have $\rho=\rhos+\rhob$.  In all the
presented cases we use $\Omegabz=\vc(\rstar)/\rstar$, where
$\vc(\rstar)$ is the circular velocity at $\rstar$ (i.e. at the edge
of the bar).

For all the simulations, the inner logarithmic slope $\gamma$, the
internal acceleration ratio $\kappa$, the total dark to luminous mass
ratio $\MDM/\Mstar$, the initial angular velocity of the bar
$\Omegabz$ and the total number of particles $\Npart$ are given in
Table~1.

\section{Results}

CB04 found that the DF timescales in a (deep) MOND system ($\tfricM$),
and in the {\it equivalent} Newtonian system ($\tfricN$) are related
by
 \begin{equation}
{\tfricM\over\tfricN}={\sqrt{2}\over 1 + \mathR},
\end{equation}
 where $1+\mathR\equiv {g/\gstarN}$ with $g$ the modulus of the MOND
field and $\gstarN$ the modulus of the Newtonian field {\it generated
by the baryonic distribution}.  We recall again that the treatment of
CB04 was carried out for a plane-parallel distribution of field
particles.  In the present case the situation is more complicate, as
$\mathR$ is not constant in the system. Thus, for each pair of
simulations with the same value of $\kappa$ and $\gamma$ (and
therefore with the same gravitational field) we assume as fiducial
value $1+\mathR=g(\rstar)/\gstarN(\rstar)$, i.e. we evaluate $\mathR$
at the edge of the bar.  Models with lower values of the parameter
$\kappa$ are in deeper-MOND regime, so $\mathR$ increases for
decreasing $\kappa$. In particular, we considered the cases
$\kappa=0.01$, $\kappa=0.1$ and $\kappa=1$, so all the models
presented have internal accelerations $\lsim \az$.

For each simulation, Fig.~\ref{figtomega} plots against time the
angular frequency of the bar $\Omegab\equiv\varphibdot$; models with
constant-density cores are on the left and those with cuspy cores are
on the right. Systems with smaller $\kappa$ and therefore more
MOND/dark-matter dominated dynamics are lower down. In each panel the
full curve shows the MOND system, and the dotted line its Newtonian
equivalent.  In every case the bar slows down more in the MOND system
than in its ENS.

\begin{figure}
\centerline{ \psfig{file=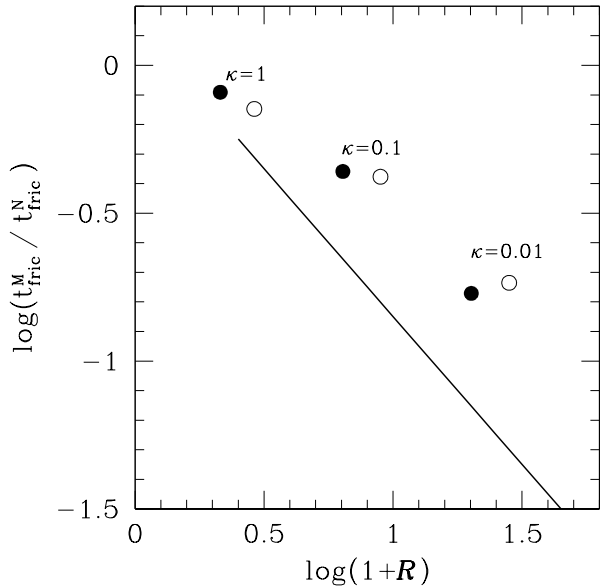,width=\hsize}}
\caption{Ratio of MOND and Newtonian equivalent DF timescale $\tfric$
  as a function of the parameter $\mathR$ for pairs of models with
  $\gamma=1$ (solid symbols) and $\gamma=0$ (empty symbols).  We
  measure $\tfric$ in the simulations as the time for $\Omegab$ to
  reach 70\% of its initial value. The solid line is the ratio as
  computed by CB04 in the deep-MOND limit.}
\label{figkratio}
\end{figure}

The instantaneous DF timescale is $|\Omegab/\Omegabdot|$, but in
practice this quantity fluctuates strongly, so is is not a useful
measure of the DF timescale. A more robust measure is the time
$\tfric$ for $\Omegab$ to reach 70\% of its initial value, so
$\Omegab(\tfric)=0.7\Omegabz$.  Figure~\ref{figkratio} is a plot
against $1+\mathR$ of the MOND value of $\tfric$ divided by $\tfric$
of the ENS. This ratio decreases with $\kappa$ as predicted by CB04,
and in the deep-MOND regime the rate of decrease with increasing
$1+\mathR$ is consistent with that predicted by CB04 (shown by the
full line).  The vertical offset between the full line and the
points could partly reflect the arbitrariness of our definition of
$\tfric$ and our assignment of values of $\mathR$ to
simulations. However, we verified that for any sensible definition of
$\mathR$, the experimental ratio $\tfricM/\tfricN$ is always closer to unity
than CB04 predicted. Presumably this finding arises because we are
considering spherical systems, while  CB04 assumed plane-parallel symmetry.

%%%%%%%%%%%%%%%%%%%% FIG 3
%%%%%%%%%%%%%%%%%%%% 

\section{Relation to other work}

\cite{TirC07a} used N-body simulations similar, in some respect, to
those reported here to compare the dynamics of galactic bars in MOND
and in equivalent Newtonian systems. They found that DF had a much
{\it smaller\/} effect on bars in MOND than in the Newtonian
cases. This finding is not in conflict with our opposed finding for
this reason: whereas our bars contain only five percent of the
galaxy's baryonic mass, in the Tiret \& Combes simulations the bars
contain the majority of the baryonic mass. Consequently, there is
significant background mass for the Tiret \& Combes bars to interact
with through DF only when dark matter is present. By contrast our bars
interact with a background that contains 95 percent of the baryonic
mass, so DF is effective.

Similar reasoning applies to the findings of Nipoti et al.~(2007c) and
\cite{TirC07b} that merging timescales are much longer in MOND than in
Newtonian gravity with dark matter: in the Newtonian case there is an
abundance of dark matter to absorb energy and angular momentum from
the stellar systems, while in MOND the energy and angular momentum has
to be absorbed by the outer parts of the stellar systems themselves.
Consequently, in the Newtonian case a lot of matter takes up a
relatively small amount of energy per particle, and the concept of DF
is applicable, while in the case of MOND a small amount of matter
takes up a lot of energy per particle, and DF is not an appropriate
concept. The prolonged merging timescale in MOND is a consequence of
the need to completely transform the orbits of the stars that are
absorbing the original orbital energy.

\section{Conclusions}
\label{secdis}

Our fully non-linear simulations of the MOND dynamics of realistic stellar
systems nicely confirm the analytical predictions of CB04. In particular,
the simulations confirm the predicted scaling of the DF
timescale with the parameter $1+\mathR$ that measures the extent to which
the actual acceleration exceeds that generated by the baryons alone in
Newtonian gravity. Consequently, the astrophysical consequences listed in
CB04 are confirmed.

Prominent among these was the implication of short DF timescales for
the existence of globular clusters in dwarf spheroidal galaxies. Even
simple Newtonian models of dwarf spheroidals and dwarf ellipticals
predict that many observed globular clusters should spiral to the
galaxy centres within a Hubble time \citep{Tremaine76,Lot01}.
Therefore the existence of objects such as the Fornax dwarf
spheroidal, which has five globular clusters but no stellar nucleus
that could be the remains of clusters that have spiralled inwards, is
a challenge to both Newtonian gravity and MOND.  In the context of the
standard cold-dark-matter theory, the survival of Fornax's globular
cluster against DF is problematic because it can be explained only if
the dark matter halo has an extended core, and not a central cusp as
predicted by the theory \citep{Goe06,SanRH06}.  From the point of view
of MOND, dwarf spheroidals are in the deep-MOND regime
\citep[e.g.][]{Ger92}, so the MOND inspiralling time of their globular
clusters is as short as the dynamical time. S\'anchez-Salcedo et
al. (2006) evaluated in detail the predictions of MOND and Newtonian
gravity for the fate of the globular clusters of the Fornax dwarf
spheroidal galaxy, and from the results of CB04 concluded that Fornax
is much more problematic for MOND than for dark matter models. This
conclusion gains strength from our confirmation of the central result
of CB04.

\section*{Acknowledgements}

We thank Scott Tremaine for useful suggestions, and the anonymous
referee for helpful comments. Some of the numerical simulations were
performed at CINECA, Bologna, with CPU time assigned under the
INAF-CINECA agreement 2007/2008.

\end{document}